\documentclass[useAMS, usenatbib]{mn2e}
\usepackage{graphicx, rotate}
\usepackage{natbib, lscape}
\usepackage{latexsym, amssymb, verbatim}

\title[Chemistry at high CR]{Chemistry in Cosmic-Ray Dominated Regions (CRDRs)}
\author[Bayet et al.]{E. Bayet$^{1}$\thanks{E-mail:
eb@star.ucl.ac.uk; daw@star.ucl.ac.uk; twh@ast.leeds.ac.uk;
sv@star.ucl.ac.uk}; D.A. Williams$^{1}$; T. W. Hartquist$^{2}$
and S. Viti$^{1}$\\
$^{1}$Department of Physics and Astronomy, University College
London, Gower Street, London WC1E 6BT, UK\\
$^{2}$School of Physics and Astronomy, University of Leeds, Leeds
LS2 9JT, UK\\}

\begin{document}

\date{Accepted ; Received ; in original form }

\pagerange{\pageref{firstpage}--\pageref{lastpage}} \pubyear{2010}

\maketitle

\label{firstpage}

\begin{abstract}
Molecular line observations may serve as diagnostics of the degree
to which the number density of cosmic ray protons, having energies
of 10s to 100s of MeVs each, is enhanced in starburst galaxies and
galaxies with active nuclei. Results, obtained with the UCL\_PDR
code, for the fractional abundances of molecules as functions of
the cosmic-ray induced ionisation rate, $\zeta$, are presented.
The aim is not to model any particular external galaxies. Rather,
it is to identify characteristics of the dependencies of molecular
abundances on $\zeta$, in part to enable the development of
suitable observational programmes for cosmic ray dominated regions
(CRDRs) which will then stimulate detailed modelling. For a number
density of hydrogen nuclei of of $10^4$ cm$^{-3}$, and high visual
extinction, the fractional abundances of some species increase as
$\zeta$ increases to $10^{-14}$ s$^{-1}$, but for much higher
values of $\zeta$ the fractional abundances of all molecular
species are significantly below their peak values. We show in
particular that OH, H$_{2}$O, H$_{3}^{+}$, H$_{3}$O$^{+}$ and
OH$^{+}$ attain large fractional abundances ($\geqslant 10^{-8}$)
for $\zeta$ as large as $10^{-12}$ s$^{-1}$. HCO$^{+}$ is a poor
tracer of CRDRs when $\zeta > 10^{-13}$ s$^{-1}$. Sulphur-bearing
species may be useful tracers of CRDRs gas in which $\zeta \sim
10^{-16}$ s$^{-1}$. Ammonia has a large fractional abundance for
$\zeta \leqslant 10^{-16}$ s$^{-1}$ and nitrogen appears in
CN-bearing species at significant levels as $\zeta$ increases,
even up to $\sim 10^{-14}$ s$^{-1}$. In this paper, we
also discuss our model predictions, comparing them to recent
detections in both galactic and extragalactic sources. We show
that they agree well, to a first approximation, with the
observational constraints.
\end{abstract}


\section{Introduction}\label{sec:intro}

The high spatial density of massive star formation in mergers and
starburst galaxies (e.g. \citealt{Acer09}; \citealt{Such93};
\citealt{Acci09}) creates regions of extremely high cosmic ray
energy density, up to about ten thousand times that in the Milky
Way Galaxy. \citet{Papa10a} has proposed that these large energy
densities alter the heating rates and ionization fractions in
dense gas ($n$(H$_{2}$) $\gtrsim 1\times 10^{4}$ cm$^{-3}$) in the
UV-shielded cores that contain much of the molecular gas in these
galaxies, so that these cosmic ray-dominated regions (CRDRs) have
different initial conditions for star formation
(\citealt{Papa10a}; \citealt{Papa10b}). These conditions affect
the subsequent evolution of the gas and may even lead to a
top-heavy Initial Mass Function and bimodal star formation
\citep{Papa10b}.

It is therefore important to determine whether there may be useful
molecular tracers of CRDRs. \citet{Papa10a} has suggested several
possible chemical signatures, but it seems necessary to make a
fairly complete and self-consistent thermal/chemical model of
dense gas subjected to very high fluxes of cosmic rays.
Historically, the cosmic ray ionization rate in Galactic diffuse
and molecular clouds has normally been determined by treating it
as a free parameter and matching the predicted chemistry to that
observed (e.g. \citealt{Blac77, Hart78a}a and \citealt{Hart78b}b).
Here, it is necessary to reverse that procedure and to compute the
variation of the chemistry as the ionization rate is increased.
\citet{Lepp98} made such a calculation and computed the chemistry
in dense gas subjected to ionization rates varying over a wide
range. However, their calculation was for a fixed temperature.
Later, \citet{Meij05b}, \citet{Spaa05}, and \citet{Meij06}
concentrated on a restricted chemical network with enhanced
ionization rate. Recently, \citet{Baye09a} compared the computed
chemistry for a large network of 131 species connected in 1700
reactions at just two particular cosmic ray ionization rates (the
canonical Milky Way value and 100 times this value) in a
self-consistent thermal/chemical and time-dependent
one-dimensional model. In the present work, we use the same model
as that of \citet{Baye09a} and compute the chemistry for many
ionization rates from Milky Way values up to about one million
times larger.

The thermal$/$chemical model is described in Section
\ref{sec:model}, and the input parameters are discussed in Section
\ref{sec:param}. The outputs of the computations are the
temperature and the chemical abundances as functions of visual
extinction (from A$_{\rm v}$ = 3 to 20 magnitudes) and of the
cosmic ray ionisation rate (from $\zeta=2 \times 10^{-17}$
s$^{-1}$ to $\zeta=5 \times 10^{-11}$ s$^{-1}$). The model outputs
are presented in Section \ref{sec:resu}. The relevance of our
modelling to observations of galaxies is considered in Section
\ref{sec:obs}, where we also give our conclusions.

\section{Model description}\label{sec:model}

The model used for our study of the chemistry in CRDRs is the
UCL\_PDR model as described in \citet{Bell05, Bell06}. This is a
time-dependent Photon-Dominated Region (PDR) model with constant
density. In the present application, the code is run for 10$^{8}$
yrs by which time chemical steady state is reached in all models.
Checks that we have made confirm that the fractional abundances at
10$^{6}$ yrs do not differ significantly from those at 10$^{8}$
yrs. However, even for high values of $\zeta$, the chemistry
requires close to $10^6$ yrs to approach steady state. Though the
timescale to convert atomic carbon to CO decreases with increasing
$\zeta$, neutral-neutral reactions limit the rates at which many
of the observable species form. Those rates do not depend
sensitively on $\zeta$ for values in excess of about $10^{-17}$
s$^{-1}$. However, they do decrease with decreasing metallicity.

The code operates in one space dimension for an assumed
semi-infinite slab geometry, and computes self-consistently the
chemistry and the temperature as functions of depth and time
within a semi-infinite slab, taking account of a wide range of
heating and cooling processes. In the present work, the code is
used to determine the chemical and thermal properties at all
depths up to 20 visual magnitudes. For the present work, and
consistent with the \citet{Baye10a} version of the code,
additional radiative cooling due to rotational transitions of
$^{13}$CO, C$^{18}$O, CS, OH and H$_2$O have been implemented. The
escape probability formalism of \citet{DeJo80} is used to
determine non-LTE level populations and resulting line intensities
at each depth- and time-step, in the same manner as for the
existing coolants in the code. Collisional rates with H$_2$ are
taken from the Leiden Atomic and Molecular Database (LAMDA;
\citealt{Scho05}). The inclusion of these coolants allows the
thermal balance to be more accurately determined in warm dense gas
at high extinction. The UCL\_PDR code also includes cooling due to
the OI metastable levels, and the O - H$^+$ collisions and O -
e$^-$ collisions which excite them; these are important for the
models for high values of $\zeta$. The treatment of H$_2$
self-shielding has also been updated to use the results of
detailed calculations performed by \citet{Lee96}, as described in
\citet{Baye10a}. The chemical network links 131 species in over
1834 gas-phase reactions; only H$_{2}$ is formed by surface
chemistry; freeze-out of species on to grain surfaces is excluded.
The UCL\_PDR code has been validated against all other commonly
used PDR codes \citep{Roel07}.

The model should be appropriate for regions with temperatures up
to several thousand degrees. However, it is not valid for regions
with temperatures approaching $10^4$ K and the associated high
fractional ionisations. Thus, we present results only for models
for which $\zeta$ is 10$^{-12}$ s$^{-1}$ or less. Codes such as
MOCASSIN \citep{Erco05} or CLOUDY \citep{Ferl98} might be more
suitable for models with higher values of $\zeta$. The model does
not include doubly ionised species other than Ca$^{++}$, though
cosmic-ray induced ionisation will produce a variety of them (see
\citealt{Shaw08}). Charge transfer with neutral hydrogen atoms
will contribute substantially to the removal of many of them in
regions with $\zeta$ ranging from 10$^{-14}$ to 10$^{-12}$
s$^{-1}$.

\section{Parameter Selection}\label{sec:param}

The model requires the setting of a number of physical and
chemical parameters. The choices we have made are listed in Table
\ref{tab:1}. We comment here on some of these. All the
calculations are made for an assumed gas number density of
$1\times 10^{4}$ hydrogen nuclei (in all forms) cm$^{-3}$.

The cosmic ray ionisation rate is a free parameter in these
calculations, and allowed to take values that range from
$\zeta=2\times 10^{-17}$ s$^{-1}$ to 1 $\times$ 10$^{-12}$
s$^{-1}$.

The lowest value of $\zeta$ selected here is comparable to that
inferred for the Milky Way, while the highest value of $\zeta$ is
comparable to the highest values suggested to be appropriate for
compact extragalactic starburst regions (cf. eqn. 6 of
\citealt{Papa10a}).

Galaxies are observed to have a range of metallicities (e.g.
\citealt{Baye06}, \citealt{Bell06}); we have considered a range
from 0.1 to 4 times solar metallicity. Most of our computations
are made using metallicity of 0.1 times solar, a value that may
represent conditions in fairly young galaxies. However
measurements of metallicity in some galaxies show both near-solar
and larger than solar values (NGC 253, M 82, IC 342 - see e.g.
\citealt{Zari94}), so we have also made calculations with higher
metallicities.

Several other parameters are assumed to scale linearly with
metallicity. These are the dust:gas mass ratio, the H$_{2}$
formation rate, and the initial elemental abundances. The values
of these parameters for solar metallicity are given in Table
\ref{tab:1}. The elemental abundances given there are derived from
\citet{Sava96, Sofi97, Meye98, Snow02}; and \citet{Knau03}.

The models require the specification of the mean dust grain radius
and albedo, which affect the transfer of external radiation into
the cloud. The values given in Table \ref{tab:1} are canonical.
The external radiation field impinging on the one-dimensional
semi-infinite slab is assumed  to be 1000 Habing, much larger than
for the Milky Way but possibly appropriate for regions of active
star formation in which CRDRs may be located. This value is
comparable to values obtained from molecular line intensity
modelling for several galaxies (e.g. \citealt{Baye06} and
\citealt{Aalt08}). The chemistry in the one-dimensional
semi-infinite slab is explored as a function of depth or visual
extinction, A$_{\rm v}$, from the edge (A$_{\rm v}=0$) up to
A$_{\rm v}=20$ mag. The model also requires for line-width
calculations the specification of the microturbulence velocity. We
use the value adopted by \citet{Baye09c}.

\section{Results}\label{sec:resu}

\begin{figure}
    \centering
    \includegraphics[width=5cm]{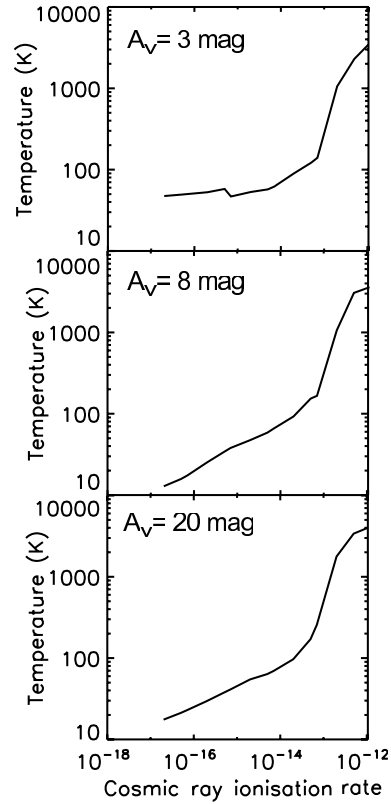}
    \caption{Temperature (in K) as a function of $\zeta$
    (in s$^{-1}$) for A$_{\rm v}=3$ mag (top), 8 mag (middle)
    and 20 mag (bottom). See the text in Sect.
    \ref{sec:resu}.}\label{fig:1}
\end{figure}

We present in Fig \ref{fig:1}-\ref{fig:4} results for the
temperature and chemistry at the three representative depths into
the one-dimensional slab: A$_{\rm v}=3$, 8 and 20 mag.

The case A$_{\rm v}=3$ is intended to represent translucent
material, A$_{\rm v}=8$ to represent the extended PDR gas (as
detected in the nuclei of M82, NGC 253, IC 342 and NGC 4038 by
\citealt{Baye09a}), while A$_{\rm v}=20$ represents dense PDR
components detected in galaxies \citep{Baye08a}.

Fig. \ref{fig:1} shows the results for temperature obtained from
the self-consistent thermal balance, as a function of cosmic ray
ionisation rate, at these three values of A$_{\rm v}$. All three
curves show the same general behaviour, rising from low values at
low ionisation rates to temperatures approaching $1\times 10^{4}$
K for the highest ionisation rates. As will be evident in the
plots of the chemical abundances, reactions at these very high
temperatures destroy much of the chemistry. There is some
difference in temperature between the three curves shown for low
values of the ionisation rate. At A$_{\rm v}=3$ mag, the
temperature lies between 50 and 100 K for ionisation rates less
than about $1\times 10^{-14}$ s$^{-1}$. This relatively high
temperature is maintained by the intense external radiation field.
However at A$_{\rm v}=8$ mag, the temperature falls to $\sim$10 K
at the lowest ionisation rates, as expected, since the external
radiation field no longer plays a role.

We note that the models give slightly higher temperatures at
A$_{\rm v}=20$ mag than at A$_{\rm v}=8$ mag. This is due to the
fact that at high visual extinctions the C and CO lines become
optically thick and hence are less able to cool the gas whilst the
cosmic ray heating remains high and constant. For some models,
there were some small difficulties associated with the convergence
of the thermal balance. For example one sees small spikes in the
thermal structure curve for A$_{\rm v}$ = 3 and corresponding
features in the plots of the fractional abundances. These
difficulties are not due to bistability \citep{Boge06}. Rather
they arise due to the stiffness of the coupled chemical and
thermal balance equations. Their influence on the calculated
abundances of observable species is not significant. The effects
of high heating rates on the chemistry is further investigated in
Bayet et al. (2010).

Fig. \ref{fig:2}, \ref{fig:3} and \ref{fig:4} present the results
giving the abundances of various relevant species at the three
specified values of A$_{\rm v}$. These species were selected from
the 131 available partly because many of them have already been
detected in external galaxies, and partly to illustrate the
sensitivity of species to the ionisation rate for a wide chemical
variety. The range of abundances shown extends somewhat beyond the
values that may be detectable.

Although there are significant differences, the three cases are
broadly similar in their behaviour with respect to the enhancement
of the ionisation rate. As the ionisation rate is increased from a
value appropriate for the Milky Way, a rich chemistry is
maintained up to a critical value of the ionisation rate, beyond
which the molecular abundances fall rapidly as the ionisation rate
is further increased. If the ionisation rate is increased to
1$\times$10$^{-12}$ s$^{-1}$, almost all the chemistry is
effectively suppressed. The driver of this decline in molecular
abundances is linked to the decline in molecular hydrogen; this
occurs when the ionisation rate is about 1$\times$10$^{-14}$
s$^{-1}$, and H$_2$ becomes a minor species when the ionisation
rate is as large as 10$^{-12}$ s$^{-1}$. In these conditions,
conventional astrochemistry - based on reactions with H$_2$
molecules - ceases. In this range of ionisation rate, the
temperature rises abruptly to some thousands of Kelvin (see Fig.
1), and hot atomic hydrogen is very destructive of molecules.
However, we can see from the figures that potential molecular
tracers can be identified for ionisation rates of about
1$\times$10$^{-13}$ s$^{-1}$ but the molecular abundances drop
substantially as $\zeta$ increases much more.

We now confine our remarks to the chemically richer regions of
parameter space. There are some differences between the panels in
Fig. \ref{fig:3} (or \ref{fig:4}). Some species decline with
increasing ionisation rate more rapidly than others. For example,
sulphur-bearing species may be useful tracers of gas in which
$\zeta = 10^{-16}$ s$^{-1}$ but not when $\zeta= 10^{-14}$
s$^{-1}$. Ammonia has a large fractional abundance for lower
ionisation rates, but the abundance declines rapidly as the
ionisation rate increases. Nitrogen appears in CN-bearing species
at levels that are probably significant as the ionisation rate
increases, even up to $\sim 10^{-14}$ s$^{-1}$.

There are some differences between the figures shown. At A$_{\rm
v}$ = 3, molecular abundances are significantly lower than those
for higher visual extinctions. For example, CS is $\sim 10^4$
times more abundant at A$_{\rm v}$ = 8 than at A$_{\rm v}$ = 3.
The intense radiation field assumed in these models to impinge on
the slab is responsible for inhibiting the chemistry at A$_{\rm
v}$ = 3. However, the model results for A$_{\rm v}$ = 8 and 20
magnitudes are essentially identical, as the external radiation
field does not penetrate effectively to these depths.

The behaviour of oxygen and carbon hydrides and their ions is
somewhat different to the above behaviours. Some of these species
show fractional abundances of rather high levels for ionisation
rates of $10^{-14}$ s$^{-1}$, while others even sustain these high
values for ionisation rates approaching $\sim 10^{-12}$ s$^{-1}$.
For example, OH and H$_2$O fractional abundances are
$\sim$10$^{-7}$ for an ionisation rate of $\sim$ 10$^{-14}$
s$^{-1}$, while OH and OH$^+$ may be as large as $\sim$10$^{-8}$
even for an ionisation rate of $\sim$ 10$^{-12}$ s$^{-1}$. The
ions H$_3^+$ and H$_3$O$^+$ are likely to be abundant for $\zeta =
10^{-13}$ s$^{-1}$; however, the commonly used tracer HCO$^+$ is
probably useless for values of $\zeta$ larger than 10$^{-13}$
s$^{-1}$. The behaviour of these oxygen and carbon species is a
consequence of the thermal stimulation of the endothermic
reactions that may initiate oxygen and carbon chemistry by the
increase in kinetic temperature arising from the higher ionisation
rates. However, ultimately, the reduction in the H$_2$ fraction
suppresses these reaction networks.

Given that the metallicity in some galaxies has been measured (see
Section \ref{sec:obs}) and the values found to lie in a range
often below but sometimes exceeding the solar value, it is
worthwhile to explore the predictions of molecular abundances in
our model for varying metallicity. Table \ref{tab:3} shows the
computed fractional abundances for some atoms and molecules of
observational interest for three values of the metallicity (0.1,
1.0 and 4.0 times the solar value), for both ``high'' ($10^{-13}$
s$^{-1}$) and ``low'' ($10^{-16}$ s$^{-1}$) cosmic ray ionisation
rates, and for two values of the visual extinction (A$_{\rm v}=$ 3
and 20 magnitudes). The results for the case when A$_{\rm v}=$ 8
mag are closely similar to those for A$_{\rm v}=$ 20 mag, and are
not shown. The ``high'' ionisation rate is intended to represent
the case predicted by \citet{Papa10a} for CRDRs, while the ``low''
case is expected to be roughly appropriate for galaxies similar to
the Milky Way.

As has been found by earlier studies of the dependence of
chemistry on the metallicity, the behaviour is rather complicated.
At first sight one may expect a higher metallicity to lead to
larger fractional abundances of molecules. However, because of the
chemical rate equations are strongly coupled, some species tend to
increase with metallicity, while others decrease, for a given
ionisation rate. Further, Table \ref{tab:3} shows that different
behaviours may occur when the ionisation rate is changed. For
example, the fractional abundance of C$_{2}$H at A$_{\rm v}=$ 3
mag with a high ionisation rate is a strongly increasing function
of metallicity, while for A$_{\rm v}=$ 20 mag and a low ionisation
rate the C$_{2}$H fractional abundance declines rapidly with
increasing metallicity.

CO is generally fairly strongly dependent on metallicity as it is
formed in successive reactions. However, at A$_{\rm v}=$ 3 mag
with a low metallicity, the molecule does not attain an abundance
high enough to provide self-shielding against the radiation field
which destroys this molecule through line absorption. Thus, at low
metallicity, the fractional abundance of CO is also low for
A$_{\rm v}=$ 3 mag. Consequently, HCO$^{+}$ is similarly affected.

These kinds of behaviour can be understood in terms of the rates
of the various competing processes. At A$_{\rm v}=$ 3 mag, the
intense radiation field adopted in these calculations (1000
Habing) may lead to photodissociation being a competitive process
with ion-molecule reactions, while at higher values of A$_{\rm v}$
photodissociation is negligible. The ionisation rate drives the
chemistry when photo-processes do not play a role, and high
ionisation rates drive the chemistry much faster than low rates.
Thus, in the case of C$_{2}$H, the dependence on metallicity is
strong because it is formed by successive reactions involving
C$^+$, itself dependent on metallicity. Thus, at A$_{\rm v}=$ 3
mag both photons and high fluxes of cosmic rays drive the
chemistry strongly to create large abundances at high
metallicities. However, at A$_{\rm v}=$ 20 mag with low ionisation
rates, the C$^{+}$ and C$_{2}$H abundances never attain high
values, even at the highest metallicity. It is evident, therefore,
that the choice of potential tracers depends sensitively on
ionisation rate as well as metallicity.

\section{Discussion and Conclusions}\label{sec:obs}

In this section we briefly explore the relevance of our
theoretical predictions for galactic and extragalactic
environments. We remind the reader that our attempt in
this paper is not to model any particular source but rather to
indicate qualitatively if our model predictions (i.e. the
variations of the species fractional abundances with respect to
the changes in cosmic ray ionisation rate) agree to a first
approximation with observations in the local and nearby Universe.

\begin{figure*}
    \centering
    \includegraphics[width=14cm]{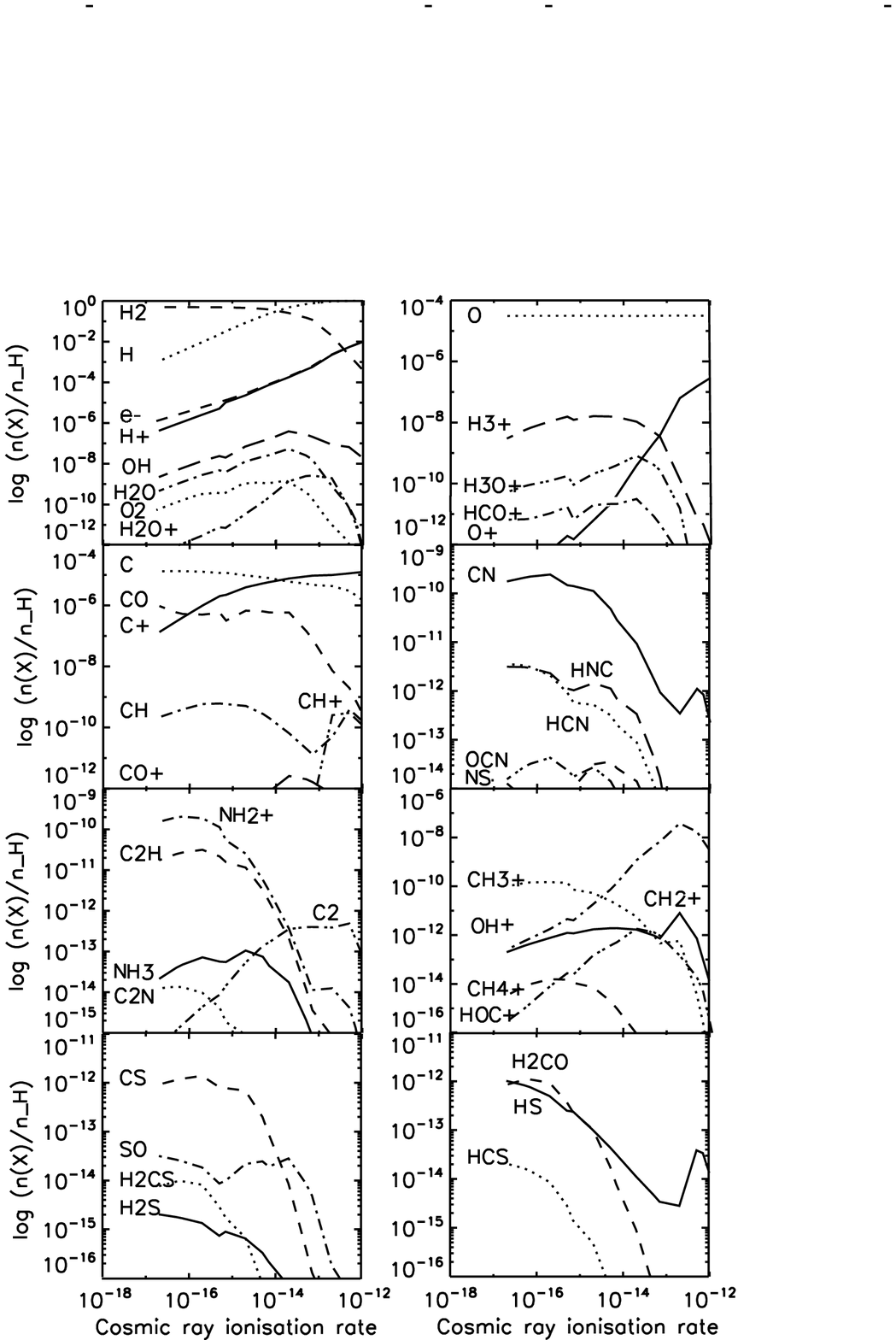}
    \caption{Fractional abundances (n(X)/n$_{\rm H}$)
    of species X in logarithmic scale where n$_{\rm H}$ is the
    total number of hydrogen atoms, obtained for the models with
    a metallicity of 0.1 z$_{\odot}$ for an A$_{\rm v}$
    of 3 mag as functions of cosmic ray ionisation rate ($\zeta$) in s$^{-1}$ (see text in Sect. \ref{sec:resu}).}\label{fig:2}
\end{figure*}

\begin{figure*}
    \centering
    \includegraphics[width=14cm]{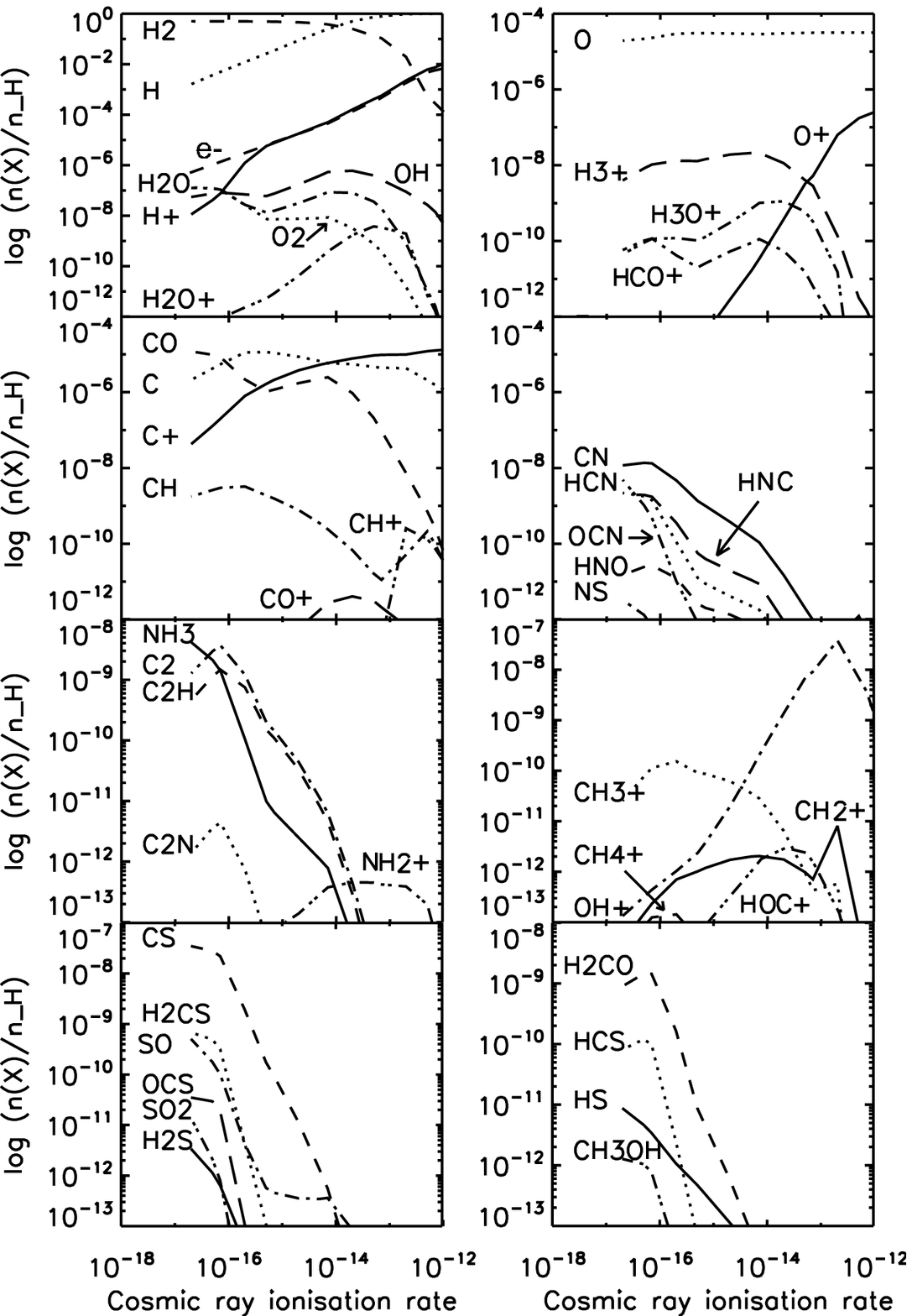}
    \caption{Fractional abundances of species X obtained for the models with a
    metallicity of 0.1 z$_{\odot}$ for an A$_{\rm v}$ of 8 mag
    (see text in Sect. \ref{sec:resu} and the caption of Fig. \ref{fig:2}).}\label{fig:3}
\end{figure*}

\begin{figure*}
    \centering
    \includegraphics[width=14cm]{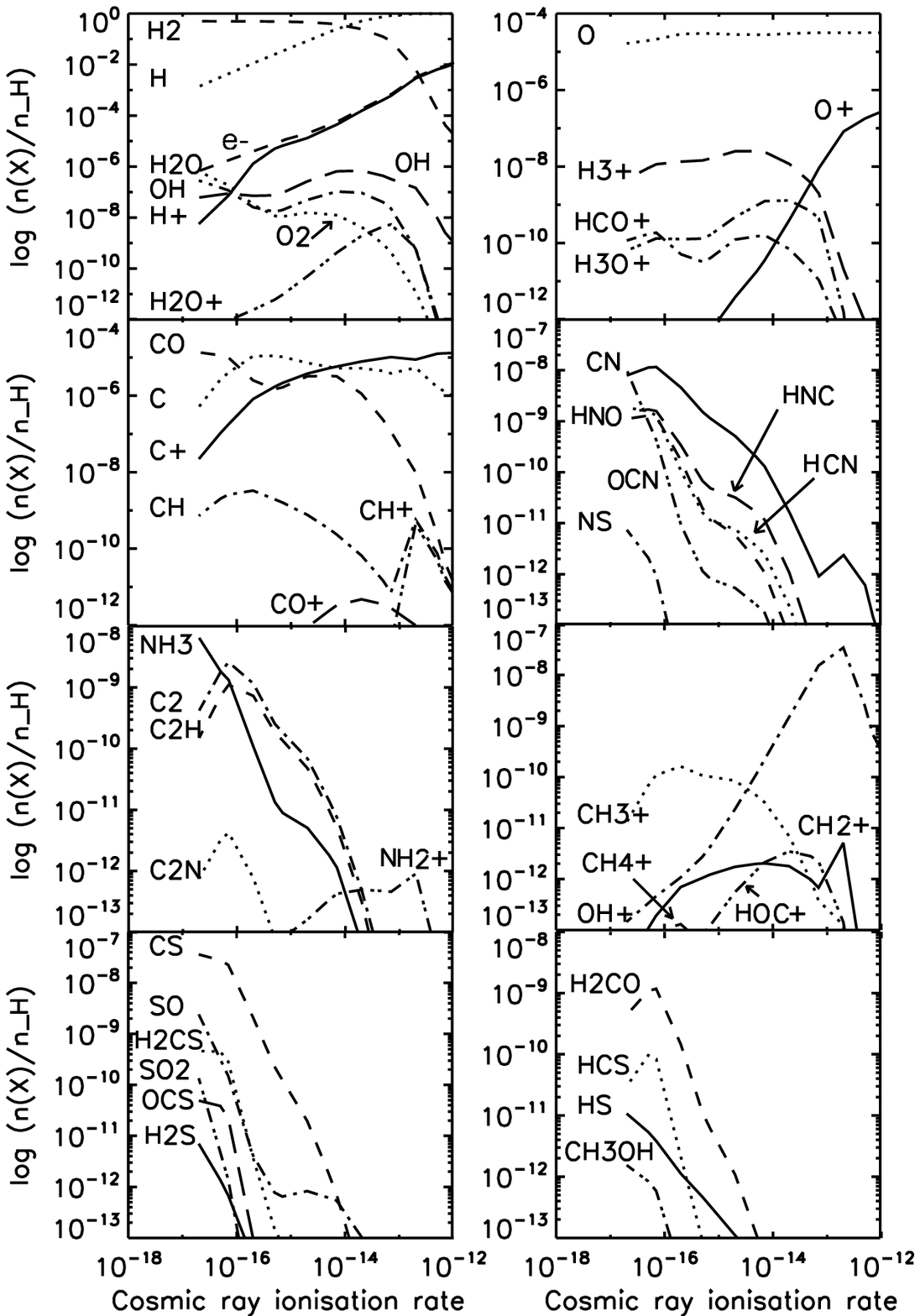}
    \caption{Fractional abundances of species X obtained for the models with a
    metallicity of 0.1 z$_{\odot}$ for an A$_{\rm v}$ of 20
    mag (see text in Sect. \ref{sec:resu} and the caption of Fig. \ref{fig:2}).}\label{fig:4}
\end{figure*}

Recent results indicate that higher cosmic ray ionization
rates might be present in diffuse clouds throughout the Galaxy and
in the Orion bar (e.g. \citealt{Indr07, Shaw08, Indr09} and even
more recent results from Herschel/HIFI such as \citealt{Gupt10,
Neuf10}). More precisely, \citet{Indr07} presents H$_{3}^{+}$
column densities varying between $0.2-2.1 \times 10^{14}$
cm$^{-2}$ depending on the diffuse cloud observed (see their
Tables 3 and 4). They derive estimates of $\zeta \sim 0.5-3.2
\times 10^{-16}$s$^{-1}$. If we convert these column densities
into fractional abundance estimates (i.e. $4.4 \times 10^{-9} -
4.4 \times 10^{-8}$) and compare them with those provided by our
model showing the smallest A$_{\rm v}$ (i.e. Fig. \ref{fig:2} for
A$_{\rm v}=3$ mag) and $\zeta=2 \times 10^{-16}$s$^{-1}$, we find
a very good agreement. Despite this encouraging agreement, the
interpretation of the H$_{3}^{+}$ observations is a highly debated
subject and a more detailed analysis, requiring the modelling of a
particular set of galactic sources (outside the scope of this
paper) is required for any definitive conclusions. \citet{Indr09}
invoke a low-energy ($\sim$ 10 MeV) cosmic ray flux for a
plausible explanation for the observed H$_{3}^{+}$ results.
Chemical models such as ours cannot be used to infer details of
the cosmic ray spectrum responsible for inducing the ionization.

Towards W49N, \citet{Neuf10} estimated column densities of
OH$^{+}$ varying between $2.2 \times 10^{13}$ cm$^{-2}$ and $2.6
\times 10^{14}$ cm$^{-2}$ whereas column densities of
H$_{2}$O$^{+}$ range from $4.2 \times 10^{12}$ cm$^{-2}$ to $2.7
\times 10^{13}$ cm$^{-2}$. These authors also inferred that these
species are located in clouds of low molecular fraction. From
their column densities we derive fractional abundances estimates
of OH$^{+}$ of about $4.5\times 10^{-9}-5.4\times 10^{-8}$ and for
H$_{2}$O$^{+}$ of about $8.8\times 10^{-10}-5.6\times 10^{-9}$.
OH$^{+}$ fractional abundances are in agreement with our models
having $\zeta>5 \times 10^{-14}$s$^{-1}$ (see Fig. \ref{fig:2})
whereas H$_{2}$O$^{+}$ fractional abundances are closer to
predictions from models having $\zeta \geqslant 10^{-14}$s$^{-1}$
(see also Fig. \ref{fig:2}). For values of $\zeta$ in this range,
Figure \ref{fig:2} also shows that $n$(H)/$n$(H$_{2}$) becomes
large, so that the gas is mainly atomic. This is consistent with
the inference of \citet{Neuf10} that the gas has a low molecular
fraction. \citet{Geri10} detected OH$^{+}$, H$_{2}$O$^{+}$  and
H$_{3}$O$^{+}$ towards the massive star-forming region G10.6-0.4
and estimated column densities of $2.5 \times 10^{14}$ cm$^{-2}$,
$6.0 \times 10^{13}$ cm$^{-2}$ and $4.0 \times 10^{13}$ cm$^{-2}$,
respectively. The OH$^{+}$ and H$_{2}$O$^{+}$ column densities are
roughly similar to those found by \citet{Neuf10} in W49N, which
suggests that the H$_{3}$O$^{+}$ fractional abundance in G10.6-0.4
may be of the order of $\sim 10^{-9}$. Figure \ref{fig:2} shows
that values of this order can be achieved when $\zeta$ is $\sim 2
\times 10^{-14}$s$^{-1}$.

Towards Orion-KL \citet{Gupt10} showed that despite lower
OH$^{+}$ and H$_{2}$O$^{+}$ column densities than in the W49N
case, a high cosmic ray ionisation rate (i.e. $\zeta \geqslant 1-2
\times 10^{-14}$s$^{-1}$) is still required for reproducing these
observational values. Here, we found that models with $\zeta \sim
10^{-14}$s$^{-1}$ for OH$^{+}$ and models with $\zeta \geqslant 2
\times 10^{-14}$s$^{-1}$ agree with those observational column
densities. These results also seem consistent, to a first
approximation, with those of \citet{Shaw08} who showed that $\zeta
> 40 \times$ the standard value (i.e. $\zeta \sim 7 \times
10^{-14}$s$^{-1}$) is required to reproduce the abundances of
several species in $\zeta$ Persei.

Interstellar molecular oxygen has been a target of SWAS,
Odin, and - more recently - Herschel HIFI (See the talk given by
P. Goldsmith at the Stormy Cosmos meeting:
http://www.ipac.caltech.edu/ism2010/talks/Goldsmith
\_StormyCosmos2010.pdf). The Herschel results confirm and extend
the earlier results. Most galactic sources show no detectable
emission, suggesting that in regions of modest temperature the
O$_{2}$ fractional abundance is low, with limits between a few
times $10^{-9}$ to a few times $10^{-8}$. Model results reported
here show that at relatively low A$_{\rm v}$ (Fig. \ref{fig:2})
the O$_{2}$ fractional abundance peaks at $\sim 10^{-9}$ when
$\zeta \sim 1 \times 10^{-14}$s$^{-1}$. At larger A$_{\rm v}$,
then X(O$_{2}$) is $\sim 10^{-8}$ when $\zeta \sim 1 \times
10^{-16}$s$^{-1}$, but falls rapidly when $\zeta$ is larger than
$\sim 1 \times 10^{-14}$s$^{-1}$. The model results appear to be
consistent with the observational data.

We consider then in particular for the extragalactic comparison,
galaxies where AGN - and/or starburst - activities may have
enhanced the cosmic ray ionization rates as well as, in some
cases, created ULIRGs. In \citet{Baye09a} we selected well-known
galaxies, such as Arp 220 or M 82 as examples of sources with
active nuclei, and where therefore the cosmic ray ionization rate
may be enhanced. Arp 220 is the prototypical ultraluminous galaxy
while M82 is the prototypical starburst. Recently
\citep{VanderTak08} H$_{3}$O$^+$ has been discovered in these two
regions and its fractional abundance has been estimated to be
$\sim$ 2--10$\times$10$^{-9}$. These authors found that
observations of M 82 are matched by a high-$\zeta$ PDR, i.e., an
evolved starburst while X-ray models are best at reproducing the
observations of Arp 220. In fact, our models indicate that one can
obtain high abundance of this ion at low (i.e in a PDR) as well as
high (i.e dense star forming gas) extinction as long as the
$\zeta$ is $\sim$ 10$^{-13}$ s$^{-1}$ and the metallicity is
solar. At high extinction, which could represent the nuclear part
of the galaxy, the abundance is higher, providing possibly a
better match for the observations. We note here that we are not
attempting to model Arp 220 that, since as \citet{VanderTak08}
pointed out, it has a quite unusual geometry.

Another interesting object which has been recently studied in
molecular emission is Mrk 231, a ULIRG. A high resolution SPIRE
FTS spectrum reveals the presence of ions such as OH$^+$, CH$^+$
and H$_2$O$^+$ \citep{VanderWerf10}. While abundances are not
derived, we can use our Table \ref{tab:3} to determine what type
of model is able to produce high fractional abundances ($\ge$
10$^{-10}$) of these three ions. We find that the only regime is
an environment with low metallicity (0.1 solar), very high cosmic
ray ionisation rates ($\ge$ 10$^{-16}$ s$^{-1}$) and low visual
extinction. \citet{VanderWerf10} explained the high abundances of
these ions by involving XDR-chemistry. We find that, in agreement
with \citet{Papa10a}, determining the origin of molecular emission
from ULIRGs such as Mrk 231 is not trivial when both sources of
energy (CR and X-rays) are present.

Finally, our results are necessarily indicative rather than
specific in that we do not estimate molecular line intensities.
Our study was motivated by the recent investigation of filaments
around the central galaxies of clusters of galaxies by
\citet{Baye10a} and by the considerations of the effects of high
cosmic ray ionization rates in ULIRGs by \citet{Papa10a} as well
as recent Herschel results. We find that several species, many
detected in extragalactic environments, are in fact tracers of
very high ionization fractions.

The general conclusions that we can draw from this study are as
follows:

\begin{itemize}
\item The dense gas kinetic temperature in galaxies with high
cosmic ray ionisation rates is elevated above Milky Way values. It
reaches values $\ge$ 3000 K for ionisation rates of about
10$^{-12}$ s$^{-1}$; \item many potential molecular tracers can be
identified for dense gas in external galaxies in which the cosmic
ray ionisation rates are enhanced, even up to rates as large as
10$^{-13}$ s$^{-1}$; \item for galaxies with ionisation rates as
large as 10$^{-12}$ s$^{-1}$, gas of number density of 10$^4$
cm$^{-3}$ is largely neutral and atomic, with a minor component of
ions. Potential tracers of such gas are rare, but include
molecular tracers OH and OH$^+$ and atomic tracers C and C$^+$;
\item Model results for the abundances of various
molecular ions detected recently in the Milky Way and in external
galaxies appear to be reasonably consistent with an origin in
regions of enhanced cosmic ray flux; \item if CRDRs are also
regions of intense radiation fields, then translucent regions are
chemically poor relative to regions that are more optically thick;
\item for regions with A$_{\rm v}\gtrsim$ 8 mags, the chemistry
appears to be independent of depth; \item the non-linear
dependence of chemistry on metallicity may allow determinations of
metallicity through molecular observations.
\end{itemize}

\begin{table}
    \caption{Model parameters (see text in Sect. \ref{sec:param}).}\label{tab:1}
    \begin{tabular}{l r}
    \hline
    Gas Density & 10$^{4}$ cm$^{-3}$\\
    Cosmic ray ionisation rate $\zeta$ & 5$\times 10^{-11}$-2$\times 10^{-17}$ s$^{-1}$\\
    Metallicity & 0.1, 1 and 4 z$_{\odot}$\\
    Gas:dust mass ratio$^{a}$ & 100 \\
    H$_{2}$ formation rate $^{a}$  & 3$\times 10^{-18}\sqrt{T} \exp(\frac{-T}{1000})$n(H)n$_{\rm H}$ cm$^{-3}$s$^{-1}$ \\
    C/H$^{a}$ & 1.42 $\times 10^{-4}$\\
    O/H$^{a}$ & 3.20 $\times 10^{-4}$\\
    S/H$^{a}$ & 1.43 $\times 10^{-6}$\\
    N/H$^{a}$ & 6.50 $\times 10^{-5}$\\
    Mg/H$^{a}$ & 5.10 $\times 10^{-6}$\\
    He/H$^{a}$ & 7.50 $\times 10^{-2}$\\
    Si/H$^{a}$ & 8.21 $\times 10^{-7}$\\
    Fe/H$^{a}$ & 3.60 $\times 10^{-7}$\\
    Cl/H $^{a}$& 1.10 $\times 10^{-7}$\\
    Na/H$^{a}$ & 8.84 $\times 10^{-7}$\\
    Ca/H$^{a}$ & 5.72 $\times 10^{-10}$\\
    Grain size & 0.1 $\micron$\\
    Grain albedo & 0.7 \\
    External FUV radiation & 1000 Habing$^{b}$ \\
    intensity &\\
    A$_{\rm v}$ maximum & 20 mag\\
    Microturbulence velocity & 1.5 kms$^{-1}$\\
    \hline
    \end{tabular}

    $^{a}$ : z = 1 = z$_{\odot}$
    corresponds to solar values of the initial elemental abundance
    ratios while z = 1/10 z$_{\odot}$ means
    that the solar values of the initial elemental abundance ratios
    have been all divided by the same factor (of 10 in this example).
    The initial elemental abundance ratios are from \citet{Sava96,
    Sofi97, Meye98, Snow02, Knau03}. Similar assumptions are
    supposed for the H$_{2}$ formation rate coefficient and the
    dust:gas mass ratio; $^{b}$: The unit of the standard Interstellar Radiation Field
    (ISRF) intensity is a mean intensity of 1.6$\times 10^{-3}$erg
    cm$^{-2}$s$^{-1}$ integrated over 912-2400 angstroms \citep{Habi68}.
\end{table}

\begin{landscape}
\begin{table}
    \caption{Fractional abundances (n(X)/n$_{\rm H}$) of species X (see col. 1), where
    n$_{\rm H}$ is the total number of hydrogen atoms, for various metallicities and as
    a function of high cosmic ray ionisation rate ($\zeta=1 \times 10^{-13}$s$^{-1}$)
    and lower cosmic ray ionisation rate ($\zeta=1 \times 10^{-16}$s$^{-1}$). The
    fractional abundances are given as a(b) which represents a$\times 10^{\rm b}$.
    The values listed are those obtained for opacities of A$_{\rm v}=3$ mag and
    A$_{\rm v}=20$ mag (see text in Sect. \ref{sec:resu}).}\label{tab:3}
    \begin{center}
    \begin{tabular}{|r c c c c c c||c c c c c c|}
    \hline
    & A$_{\rm v}$=3 mag & & & & & & A$_{\rm v}$=20 mag & & & &\\
    $\zeta$ (s$^{-1}$)& $ 10^{-13}$ & $10^{-13}$ & $10^{-13}$ & $ 10^{-16}$ & $ 10^{-16}$ & $ 10^{-16}$
                         & $ 10^{-13}$ & $10^{-13}$ & $ 10^{-13}$& $ 10^{-16}$ & $ 10^{-16}$ &$10^{-16}$\\
    metallicity (z$_{\odot}$)& 0.1 & 1 & 4 & 0.1 & 1 & 4 & 0.1 & 1 & 4 & 0.1 & 1 & 4\\
    \hline
    Molecule &  &  & & &  & &  &  \\
    \hline
    H$_{2}$       & 2.69(-3) &3.35(-1)  & 4.50(-1)  & 1.96(-2) &4.99(-1)  & 4.99(-1)  & 4.94(-1) &3.38(-1)  & 4.53(-1)  & 4.95(-1)&4.99(-1)  & 4.99(-1)\\
    OH            & 1.11(-7) &1.19(-6)  & 4.26(-7)  & 8.75(-8)&5.66(-9)  & 3.08(-9)  & 7.20(-8)&3.63(-6)  & 6.61(-7)  & 1.37(-8)&7.53(-8)  & 3.82(-8)\\
    H$_{2}$O      & 2.64(-10)&1.75(-7)  & 9.36(-8)  & 8.06(-10)&1.64(-9)  & 1.44(-9)  & 2.55(-8)&6.12(-7)  & 1.65(-7)  & 2.80(-9)&1.94(-6)  & 5.75(-6)\\
    H$_{2}$O$^{+}$& 2.03(-10)&4.25(-9)  & 1.34(-9)  & 1.96(-9)&1.53(-12) & 1.37(-12) & 2.41(-12)&7.36(-9)  & 1.78(-9)  & 2.93(-12)&1.85(-12) & 2.69(-12)\\
    C$^{+}$       & 9.43(-6) &9.95(-5)  & 3.59(-4)  & 9.91(-6)&3.14(-6)  & 4.03(-6)  & 8.18(-7)&5.74(-5)  & 3.18(-4)  & 9.94(-7)&1.66(-8)  & 5.03(-9)\\
    C             & 4.76(-6) &3.06(-5)  & 1.68(-4)  & 4.28(-6)&8.49(-5)  & 1.33(-4)  & 1.08(-5)&1.62(-5)  & 1.24(-4)  & 1.27(-5)&6.85(-7)  & 9.11(-7)\\
    CO            & 7.16(-9) &1.19(-5)  & 4.07(-5)  & 7.27(-9)&5.40(-5)  & 4.32(-4)  & 2.59(-6)&6.84(-5)  & 1.26(-4)  & 4.99(-7)&1.41(-4)  & 5.63(-4)\\
    CH$^{+}$      & 3.82(-10)&6.37(-12) & 3.29(-11) & 2.54(-10)&5.50(-13) & 2.62(-13) & 5.03(-13)&5.98(-12) & 3.14(-11) & 5.13(-13)&8.94(-15) & 4.67(-15)\\
    CH            & 9.60(-10)&1.54(-9)  & 1.87(-8)  & 4.97(-11)&1.40(-9)  & 1.01(-9)  & 3.30(-9)&1.16(-9)  & 2.35(-8)  & 5.88(-10)&3.34(-10) & 2.77(-11)\\
    CO$^{+}$      & 8.12(-13)&9.55(-11) & 1.19(-10) & 6.64(-13)&1.86(-14) & 2.93(-14) & 6.04(-14)&1.83(-10) & 1.75(-10) & 1.18(-14)&4.97(-14) & 7.49(-14)\\
    C$_{2}$H      & 5.74(-16)&2.87(-11) & 1.15(-9)  & 8.30(-16) &2.69(-10) & 2.09(-10) & 7.28(-10)&4.73(-11) & 1.67(-9)  & 3.15(-11)&2.12(-12) & 8.35(-14)\\
    NH$_{3}$      & 9.31(-17)&3.91(-13) & 1.01(-12) & 2.03(-16) &1.71(-13) & 9.29(-14) & 1.09(-10)&4.27(-12) & 2.78(-12) & 7.18(-14)&7.93(-9)  & 4.27(-9)\\
    CS            & 2.98(-17)&1.53(-12) & 3.41(-10) & 1.55(-17)&5.20(-11) & 1.13(-10) & 2.16(-9)&7.01(-12) & 1.63(-9)  & 1.39(-12)&3.55(-7)  & 1.37(-6)\\
    SO            & 1.68(-16)&1.02(-12) & 3.31(-12) & 1.80(-16)&2.60(-12) & 2.39(-11) & 4.21(-12)&1.27(-11) & 7.34(-12) & 1.85(-14)&4.70(-8)  & 1.15(-7)\\
    H$_{2}$CS     & 7.17(-24)&1.04(-16) & 6.39(-15) & 1.45(-23)&8.40(-14) & 9.56(-14) & 3.77(-12)&5.64(-16) & 1.48(-14) & 8.14(-15)&7.68(-11) & 3.06(-11)\\
    H$_{2}$S      & 2.00(-18)&1.19(-15) & 1.15(-14) & 5.74(-19)&2.09(-14) & 8.42(-14) & 3.86(-14)&1.67(-15) & 1.53(-14) & 1.34(-15)&1.20(-11) & 1.86(-11)\\
    H$_{3}$O$^{+}$& 1.90(-13)&1.89(-9)  & 5.51(-10) & 1.53(-11)&2.45(-11) & 6.90(-12) & 1.27(-10)&6.79(-9)  & 9.29(-10) & 1.24(-10)&8.97(-11) & 7.61(-11)\\
    HCO$^{+}$     & 1.73(-13)&6.76(-10) & 7.57(-10) & 4.77(-13)&2.79(-11) & 2.02(-11) & 5.26(-11)&3.73(-9)  & 1.83(-9)  & 1.05(-11)&1.64(-10) & 1.16(-10)\\
    CN            & 6.21(-12)&6.16(-10) & 8.10(-9)  & 3.47(-13)&2.23(-9)  & 1.82(-9)  & 4.55(-9)&1.20(-9)  & 1.06(-8)  & 2.42(-10)&7.26(-9)  & 3.28(-9)\\
    HCN           & 6.85(-14)&1.88(-11) & 6.25(-10) & 3.09(-15)&2.42(-10) & 7.91(-10) & 1.06(-10)&9.07(-11) & 1.59(-9)  & 2.07(-12)&2.57(-9)  & 8.64(-9)\\
    HNC           & 6.97(-18)&5.15(-11) & 3.09(-10) & 7.68(-17)&1.82(-10) & 5.54(-10) & 3.36(-10)&2.28(-10) & 1.89(-10) & 2.31(-12)&1.87(-9)  & 2.13(-9)\\
    OH$^{+}$      & 2.52(-8)&4.42(-9)  & 1.64(-9)  & 3.77(-8)&9.93(-13) & 9.41(-13) & 1.08(-12)&5.34(-9)  & 1.92(-9)  & 1.78(-12)&1.12(-12) & 1.68(-12)\\
    HOC$^{+}$     & 8.76(-14)&5.39(-11) & 7.43(-11) & 1.46(-13)&1.80(-14) & 2.49(-14) & 5.95(-14)&1.30(-10) & 1.28(-10) & 7.98(-15)&1.10(-13) & 1.12(-13)\\
    H$_{2}$CO     & 3.53(-19)&7.21(-14) & 1.23(-12) & 1.709(-18)&4.65(-12) & 2.48(-12) & 1.43(-10)&2.08(-13) & 1.87(-12) & 9.33(-13)&1.22(-11) & 1.32(-12)\\
   \hline
    \end{tabular}
    \end{center}
\end{table}
\end{landscape}

\section*{Acknowledgments}

EB acknowledges financial support from STFC. We thank the referee
for constructive comments that have improved the original version
of the paper.

\newcommand{\apj}[1]{ApJ, }
\newcommand{\apss}[1]{Ap\&SS, }
\newcommand{\aj}[1]{Aj, }
\newcommand{\apjs}[1]{ApJS, }
\newcommand{\apjl}[1]{ApJ Letter, }
\newcommand{\aap}[1]{A\&A, }
\newcommand{\aaps}[1]{A\&A Suppl. Series, }
\newcommand{\araa}[1]{Annu. Rev. A\&A, }
\newcommand{\aaas}[1]{A\&AS, }
\newcommand{\bain}[1]{Bul. of the Astron. Inst. of the Netherland,}
\newcommand{\mnras}[1]{MNRAS, }
\newcommand{\nat}[1]{Nature, }
\newcommand{\araaa}[1]{ARA\&A, }
\newcommand{\planss}[1]{Planet Space Sci., }
\newcommand{\jrasc}[1]{Jr\&sci, }
\newcommand{\pasj}[1]{PASJ, }
\newcommand{\pasp}[1]{PASP, }

\bibliographystyle{mn2e}
\bibliography{references}

\bsp

\label{lastpage}

\end{document}